\documentclass[aip,jap,overcite,endfloats*]{revtex4-1}
\usepackage[pdftex,colorlinks=true,allcolors=blue]{hyperref}
\usepackage{graphicx}
\usepackage{dcolumn}
\usepackage{bm}
\usepackage{xcolor}
\usepackage{amsmath}
\usepackage{accents}

\usepackage{endnotes}
\let\footnote=\endnote

\begin{document}

\title{Measurement of spin mixing conductance in Ni$_{81}$Fe$_{19}$/$\alpha$-W and Ni$_{81}$Fe$_{19}$/$\beta$-W heterostrucutures via ferromagnetic resonance}

\author{W. Cao}\email{wc2476@columbia.edu}
\affiliation{Materials Science and Engineering, Department of Applied Physics and Applied Mathematics, Columbia University, New York, New York 10027, USA}
\author{J. Liu}
\affiliation{Materials Science and Engineering, Department of Applied Physics and Applied Mathematics, Columbia University, New York, New York 10027, USA}
\author{A. Zangiabadi}
\affiliation{Materials Science and Engineering, Department of Applied Physics and Applied Mathematics, Columbia University, New York, New York 10027, USA}
\author{K. Barmak}
\affiliation{Materials Science and Engineering, Department of Applied Physics and Applied Mathematics, Columbia University, New York, New York 10027, USA}
\author{W. E. Bailey}\email{web54@columbia.edu}
\affiliation{Materials Science and Engineering, Department of Applied Physics and Applied Mathematics, Columbia University, New York, New York 10027, USA}

\begin{abstract}
We present measurements of interfacial Gilbert damping due to the spin pumping effect in Ni$_{81}$Fe$_{19}$/W heterostructures. Measurements were compared for heterostructures in which the crystallographic phase of W, either $\alpha$(bcc)-W or $\beta$(A15)-W, was enriched through deposition conditions and characterized using X-ray diffraction (XRD) and high-resolution cross-sectional transmission electron microscopy (HR-XTEM). Single phase Ni$_{81}$Fe$_{19}$/$\alpha$-W heterostructures could be realized, but heterostructures with $\beta$-W were realized as mixed $\alpha$-$\beta$ phase. The spin mixing conductances (SMC) for W at interfaces with Ni$_{81}$Fe$_{19}$ were found to be significantly lower than those for similarly heavy metals such as Pd and Pt, but comparable to those for Ta, and independent of enrichment in the $\beta$ phase.
\end{abstract}

\maketitle

\section{introduction}

The heavy metals Ta, W and Pt have drawn attention as charge-to-spin-current-converters using spin Hall and related effects\cite{Saitoh_2006,Liu_2011,Pai_2012,Wang_2014}. Beta phase W, $\beta$-W, with the topologically close-packed A15 structure\cite{A15bW}, possesses a ``giant'' spin Hall angle of $\theta_{SH}\approx$ 0.3--0.4\cite{Pai_2012,Hao_2015}. The spin transport properties of $\beta$-W, such as the spin Hall angle $\theta_{SH}$ and spin diffusion length $\lambda_{SD}$, have been characterized by different methods\cite{Pai_2012,Hao_2015,Liu_2015,Demasius_2016}. In these studies, the metastable $\beta$-W layers were deposited directly on the substrate, were only stable for small W thickness, and were presumably stabilized through residual water vapor or oxygen on the substrate surface; thicker W films typically revert to the stable (bcc) $\alpha$ phase.

Recently, some of us\cite{Liu_ACTA,Liu_2017,Barmak_2017} have optimized a different method to stabilize the metastable-$\beta$-phase, using the introduction of N$_2$ gas\cite{Arita_1993} while sputtering at low power. Relatively thick (over 100 nm) monophase $\beta$-W films could be stabilized this way, when deposited on glass substrates. This technique has allowed deposition of majority $\beta$ phase W for 14 nm W films on CoFeB, as CoFeB/W(14 nm), and of minority $\beta$ phase for 14 nm W films on Ni and Ni$_{81}$Fe$_{19}$ (``Py''), as Ni/W(14 nm) and Py/W(14 nm). In the present work, we have prepared both monophase Py/$\alpha$-W (here Py/``$\alpha$''-W) and mixed phase Py/($\alpha$+$\beta$)-W (here Py/``$\beta$''-W) heterostructures using our optimized sputtering technique to enrich the fraction of $\beta$-W. Crystallographic phases of W were characterized by X-ray diffraction (XRD), and high-resolution cross-sectional transmission electron microscopy (HR-XTEM); secondary structural information was provided by electrical resistivity measurements at room temperature. We note that our measurements cannot distinguish between purely metallic, A15 $\beta$-W and A15 W oxide or nitride (e.g. W$_3$O); the identity of $\beta$-W as a purely metallic phase or a compound is a longstanding controversy\cite{Hagg1954,Arita_1993}.

In ferromagnet (FM)/normal metal (NM) heterostructures, pure (chargeless) spin currents can be injected from the FM into the NM by exciting ferromagnetic resonance (FMR) in the FM layer, ``pumping'' out spin current\cite{Heinrich2001,Tserkovnyak2002}. If the spin current is absorbed in the NM layer, the influence of ``spin pumping'' can be observed through the increase in the linewidth of the resonance, proportional to frequency $\omega$ as Gilbert damping, due to the loss of angular momentum from the precessing spin system\cite{Heinrich2001,Tserkovnyak2002}. The efficiency of the spin pumping effect for a given interface is characterized through the spin mixing conductance (SMC) $g_{FM/NM}^{\uparrow\downarrow}$. The SMC is also an important parameter for the interpretation of inverse spin Hall effect (ISHE) measurements\cite{Mosendz2010,Wang_2014}, in which the spin Hall angle $\theta_{SH}$ is measured by pumping chargeless spin current into the NM by FMR and measuring spin-to-charge current conversion through the generated charge current. Measurements of spin mixing conductance for Py/$\alpha$-W and Py/$\beta$-W have not been reported previously, although some measurements have been reported for W oxide\cite{Demasius_2016}. For these measurements, the simplest way to isolate the contribution of the FM/NM interface to the damping, and thus the spin pumping effect and spin mixing conductance $g_{FM/NM}^{\uparrow\downarrow}$, is to deposit the FM on the bottom and the NM on top, so that comparison structures without the NM layer have nearly identical microstructure. The ability to deposit enriched $\beta$-W on Py rather than on an insulating substrate is thus important for the measurement of spin mixing conductance of Py/$\beta$-W. In this manuscript, we report measurements of spin mixing conductances for Py/``$\alpha$''-W and Py/``$\beta$''-W interfaces using variable-frequency, swept-field FMR, as in our previous work\cite{Caminale2016,Yi2016,Wei2019}.

\section{sample preparation}
Ultrahigh vacuum (UHV) magnetron sputtering was used to deposit substrate/Ta(5 nm)/Cu(5 nm)/Ni$_{81}$Fe$_{19}$ (Py)/W/Cu(5 nm)/Ta(5 nm) heterostructures on both oxidized Si and glass substrates at room temperature, with base pressure better than $2\times10^{-8}$ Torr. The samples consist of two thickness series in ``$\alpha$''-W and ``$\beta$''-W for a total of four series. In the first thickness series, the thickness of Py ($t_{Py}=\textrm{5 nm}$) was fixed and the thickness of W was varied, with $t_{W}=\textrm{2, 5, 10, 30 nm}$, for both ``$\alpha$''-optimized and ``$\beta$''-optimized conditions. This thickness series was used for resistivity measurements, X-ray diffraction (XRD) ($t_{W}=\textrm{10, 30 nm}$), high-resolution cross-sectional transmission electron microscope (HR-XTEM) ($t_{W}=\textrm{30 nm}$) and FMR characterization. In the second thickness series, the thickness of W ($t_{W}=\textrm{10 nm}$) was fixed and the thickness of Py was varied, with $t_{Py}=\textrm{3, 5, 10, 20 nm}$, also for both ``$\alpha$'' and ``$\beta$' conditions. This thickness series was used only for FMR characterization. The same stacks without W layers, Py(3, 5, 10, 20 nm), were deposited as reference samples for FMR measurements. One heterostructure with reverse depostion order, ``$\alpha$''-W(10 nm)/Py(5 nm), was deposited in the absence of N$_2$ gas and characterized by XRD and FMR; this was not possible for ``$\beta$''-W because the $\beta$ phase cannot be stabilized on Cu underlayers\cite{Liu_2017}.

The W layers in all samples were deposited with 10 W power, nearly constant deposition rate ($<0.1$ \AA/s), and Ar pressure of $3\times10^{-3}$ Torr. Nitrogen gas, with $1.2\times10^{-5}$ Torr pressure measured by a residual gas analyzer, was introduced to promote the growth of $\beta$ phase W\cite{Liu_2017}.

\section{structural characterization}
Crystalline phases of W in the Py/W heterostructures were characterized primarily by XRD (Section A), with supporting measurements by HR-XTEM (Section B), and finally with some indirect evidence in room-temperature electrical resistivity measurements (Section C). Our basic findings are that films deposited without N$_2$, optimized for ``$\alpha$''-W, are nearly single-phase $\alpha$ in Py/``$\alpha$''-W, while in the Py/``$\beta$''-W optimized heterostructures, deposited in the presence of N$_2$, the W layers are mixed $\alpha$+$\beta$ phase, with a roughly 50\%--50\% mixture of $\alpha$-W and $\beta$-W averaged over a 10 nm film. The phase composition within the first 5 nm of the interface may have a slightly greater fraction of $\alpha$-W, but $\beta$-W could be positively identified here as well.

\subsection{X-ray diffraction}
Both symmetric ($\theta$-$2\theta$) and grazing-incidence, fixed sample angle X-ray diffraction (XRD) scans were carried out on Py(5 nm)/W(10 nm) and Py(5 nm)/W(30 nm) heterostructures deposited on glass substrates. The scans are compared for ``$\alpha$''-W and ``$\beta$''-W depositions. Scans were recorded using Cu $K_{\alpha}$ radiation and a commercial diffractometer.

\begin{figure}
  \includegraphics[width=\columnwidth]{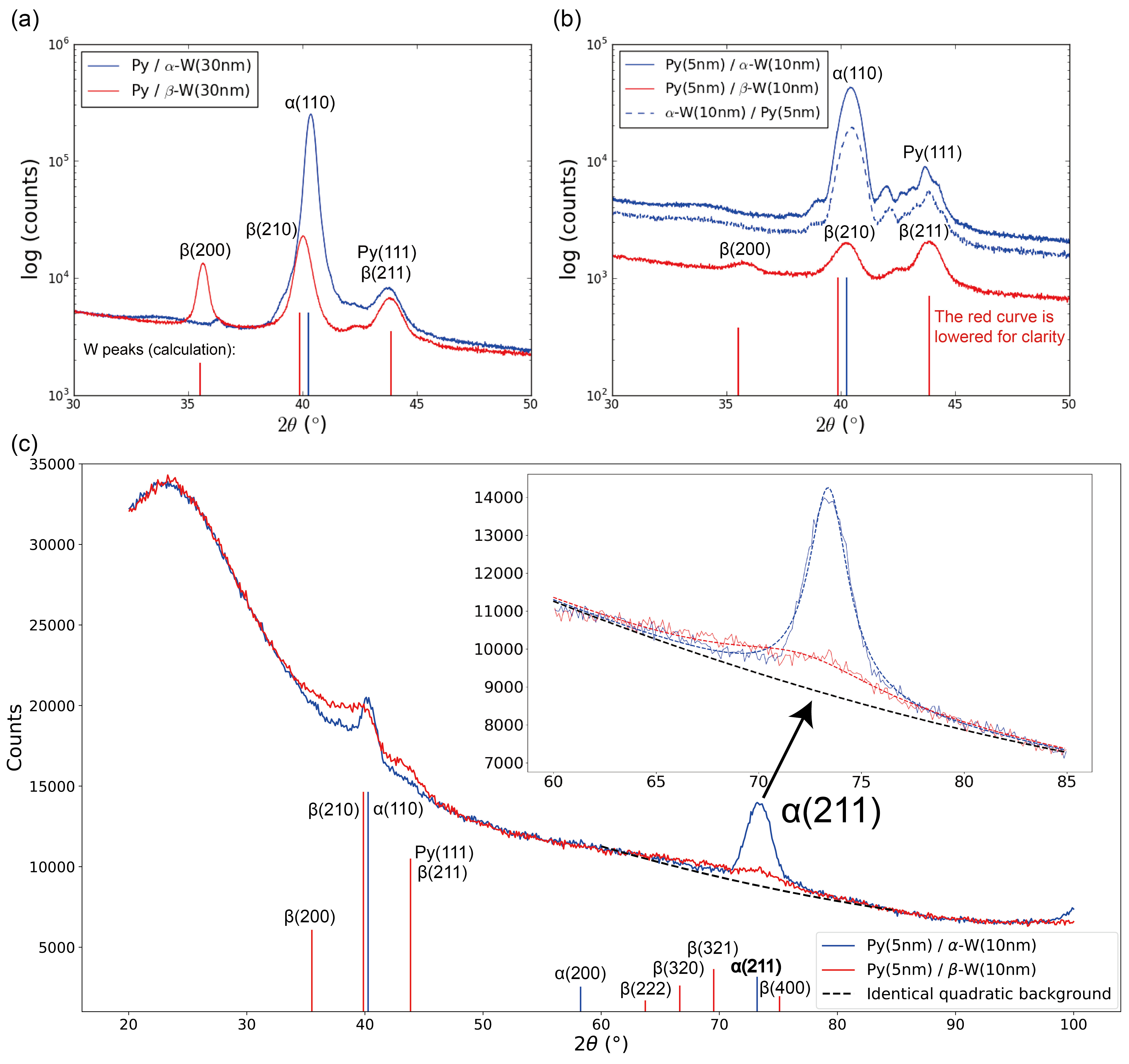}
  \caption{X-ray diffraction (XRD) measurements for Py(5 nm)/``$\alpha$''-W($t_{W}$) (blue) and Py(5 nm)/``$\beta$''-W($t_{W}$) (red) deposited on glass substrates. (a) $t_{W}=\textrm{30 nm}$; (b) $t_{W}=\textrm{10 nm}$. Solid vertical lines show the calculated reflections and intensities for $\alpha$-W and $\beta$-W peaks. (c) Grazing-incidence XRD measurements for Py(5 nm)/``$\alpha$''-W(10 nm) and Py(5 nm)/``$\beta$''-W(10 nm) samples. The inset shows the $\alpha$(211) reflections observed in both samples. The blue and the red dashed lines refer to the fits for Py/``$\alpha$''-W and Py/``$\beta$''-W, respectively. The black dashed line refers to the identical quadratic background.}
  \label{XRD}
\end{figure}

The symmetric ($\theta$-$2\theta$) scans, with scattering vector perpendicular to film planes, are presented first. We point out some obvious features of the symmetric XRD spectra, shown in Figures \ref{XRD}a) and \ref{XRD}b). For the Py/``$\alpha$''-W(30 nm) film in Fig. \ref{XRD}a), all peaks can be indexed to the close-packed planes, Cu(111)/Py(111) (fcc) and $\alpha$-W(110) (bcc). The small peak at $2\theta = 36^\circ$ can be indexed to the reflection of a small amount of Cu $K_\beta$ radiation from $\alpha$-W(110). Moving to the thinner $\alpha$ phase film in Fig. \ref{XRD}b), Py/``$\alpha$''-W(10 nm), it is still the case that all reflections can be indexed to the close-packed Cu(111)/Py(111) and $\alpha$-W(110) planes. However, there is greater structure in these reflections, presumably due to finite-size oscillations (Laue satellites), expected to be more evident in thinner films. Nearly identical spectra are recorded for the 10 nm ``$\alpha$''-W films regardless of deposition order: Py(5 nm)/``$\alpha$''-W(10 nm) and ``$\alpha$''-W(10 nm)/Py(5 nm) films scatter X-rays very similarly, as shown in Fig. \ref{XRD}b). We should note that Cu deposited on Ta has strong \{111\} texture in our films. Py (Ni$_{81}$Fe$_{19}$) deposited on Cu also has strong \{111\} texture; growth of Py on Cu and vice-versa is found to be largely coherent within grains. Both layers are fcc with similar lattice parameters: $a_{Cu}\approx 3.61$ {\AA} for Cu\cite{Liu_2017,Otte1961} and $a_{Py}\approx 3.55$ {\AA} for Py\cite{Liu_2017,HUANG_1997}, with a small misfit strain of $\epsilon=|a_{Cu}-a_{Py}|/a_{Cu}\approx 2\%$. The XRD peaks for (111)-reflections in bulk phases, broadened by finite-size effects ($FWHM\approx 1.7^\circ$ for 5 nm films, using the Scherrer equation\cite{scherrer1918,Ying_2009}), are very close to each other, at $44.2^\circ$ (Py) and $43.4^\circ$ (Cu) respectively, so we expect (and have observed) one averaged peak for Cu and Py.

The nominal ``$\beta$''-W films (red lines) clearly show the presence of the $\beta$ phase through the unique $\beta$-W(200) reflection at $2\theta \simeq 36^\circ$. This unique reflection is very strong in the ``$\beta$''-W(30 nm) heterostructure (Fig. \ref{XRD}a) but weaker as a proportion of the total intensity in the thinner ``$\beta$''-W(10 nm) heterostructure (Fig. \ref{XRD}b). In Fig. \ref{XRD}a), experimental $\beta$-W(200) and $\beta$-W(210) reflections have intensities in a ratio similar to the theoretical scattering intensity ratios for randomly-oriented $\beta$ grains. This is not the case for the thinner ``$\beta$''-W(10 nm) heterostructure in Fig. \ref{XRD}b); here the unique $\beta$-W(200) peak is less intense than expected. We interpret the relative weakness of $\beta$(200) as the presence of a large fraction of $\alpha$ grains in the nominal Py/``$\beta$''-W(10 nm) heterostructure.

In order to quantify the amount of $\alpha$-W in the nominal ``$\beta$''-W film, we have carried out grazing incidence measurements of Py(5 nm)/W(10 nm) samples ($20^\circ \leq 2\theta \leq 100^\circ$) on the same diffractometer, as illustrated in Fig. \ref{XRD}c). The samples were measured at a fixed source position of $5^\circ$ with $0.1^\circ$ step size, $0.25^\circ$ fixed slit and the 15 mm beam mask. From the TEM measurements in Fig. \ref{TEM1}b), we find that the deposited ``$\alpha$''-W films have \{110\} texture, i.e., the hexagonal arrangement ($60^\circ$ angles) of the \{011\} $\alpha$-W reflections away from the surface normal. Thus with the grazing incidence geometry, in which the scattering vector does not remain perpendicular to the film plane, the relative intensities of the peaks will not match theoretical calculations (vertical lines) based on randomly-oriented, untextured films. For example, the $\alpha$-W(200) peak (blue, $\sim 58^\circ$ in $2\theta$) almost vanishes in the XRD scan here, due to the \{011\}$\alpha$-W texture.

Here we focus on the $\alpha$-W(211) peaks near $2\theta = 72^\circ$, observed in both Py/``$\alpha$''-W and Py/``$\beta$''-W samples. As shown in the Fig. \ref{XRD}c) inset, the $\alpha$-W(211) peaks ($60^\circ \leq 2\theta \leq 85^\circ$), were fitted as the sum of Lorentzian peak and identical background, assumed quadratic in $2\theta$, for both Py/``$\alpha$''-W and Py/``$\beta$''-W samples. First we fit the $\alpha$-W(211) peak (blue) in the Py/``$\alpha$''-W sample to the summed function to determine the Lorentzian peak and quadratic background parameters. Next, we use this fitted background in the fit to the $\alpha$-W(211) peak (red) in the Py/``$\beta$''-W sample. The two fitted $\alpha$-W(211) peaks are shown as blue (for Py/``$\alpha$''-W) and red (for Py/``$\beta$''-W) dashed lines in the Fig. \ref{XRD}c) inset. The fits reproduce the experimental data well in the fitted region. The integrated $\alpha$-W(211) peak (i.e., the $2\theta$-integrated area between the measured data and the fitted background) for the Py/``$\beta$''-W sample has roughly half the intensity of the integrated peak for the Py/``$\alpha$''-W sample. Assuming that the nominal $\alpha$-W is 100\% $\alpha$ phase and that the $\alpha$ grains in mixed phase ``$\beta$''-W have similar \{110\} texture, as is supported by the HR-XTEM measurements in Figures \ref{TEM1} and \ref{TEM2}, we conclude that the Py/``$\beta$''-W(10 nm) film is roughly 50\% $\alpha$-W and 50\% $\beta$-W.

\subsection{Transmission electron microscopy}
The phases of the nominal Py(5 nm)/``$\alpha$''-W(30 nm) and the nominal Py(5 nm)/``$\beta$''-W(30 nm) samples deposited on oxidized Si substrates were characterized in high-resolution cross-sectional imaging, selected-area diffraction, and focused-beam nanodiffraction, by transmission electron microscopy (for details see the endnote\footnote{Focused ion-beam (FIB) and FEI Helios NanoLab 660 were used to prepare foils for TEM studies. To protect the heterostructures against the ion-beam damage during sample preparation, amorphous Platinum (1.5 $\mu m$ thick) was sputtered on the surface of the wafers by electron and ion beam, respectively. TEM and high-resolution cross-sectional TEM (HR-XTEM) analyses were performed by image Cs-corrected FEI Titan Themis 200 at an accelerating voltage of 200 kV. Nano-beam electron diffraction pattern (DP) technique and Fourier transform (FT) analysis of the HRTEM have been utilized to identify the nature of each phase at the scale of 1--2 nm wide. The nano-beam DPs were obtained by FEI Talos TEM operating at 200 kV. The second condenser aperture was set to 50 $\mu m$ to obtain a small beam-convergence angle. In the diffraction mode, the beam was condensed to a spot ($\sim$ 1--2 nm) and a convergent electron beam diffraction (in this case, known as Kossel-M\"{o}llenstedt pattern) was acquired at different locations on the sample.}).

\begin{figure}
  \includegraphics[width=\columnwidth]{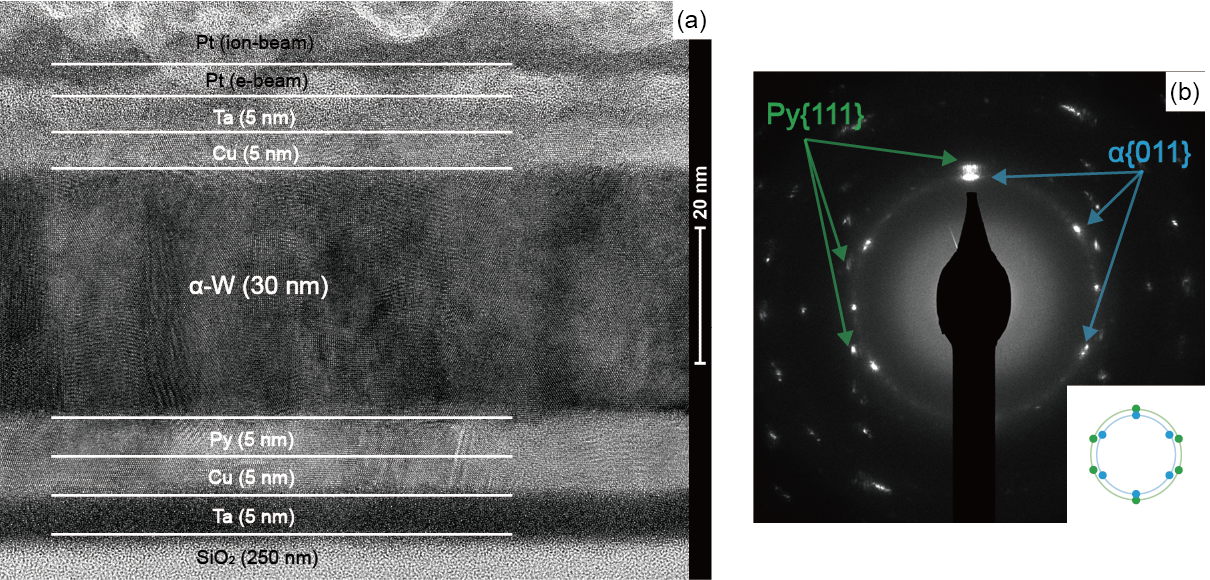}
  \caption{(a) High-resolution cross-sectional transmission electron microscopy (HR-XTEM) image of SiO$_{2}$/Ta(5 nm)/Cu(5 nm)/Py(5 nm)/``$\alpha$''-W(30 nm)/Cu(5 nm)/Ta(5 nm) heterostructure. The $\alpha$-W grains are columnar with lateral radius of 10--20 nm, with larger grain size in the growth direction. (b) Selected-area diffraction (SAD) pattern of the heterostructure, showing the preferred texture of $\alpha$-W grains on Py layer, \{111\}Py//\{011\}$\alpha$-W (see calculated pattern in the inset). No sign of $\beta$-W was detected in this heterostructure.}
  \label{TEM1}
\end{figure}

Fig. \ref{TEM1} shows a cross-sectional image and diffraction pattern for the nominal Py/``$\alpha$''-W(30 nm) heterostructure. First, one can see from the mass contrast between W and the 3d transition metal elements (Ni, Fe, Cu) that the Py/W and W/Cu interfaces are relatively flat and sharp on the scale of the image resolution of $\sim 3$ nm, presumably broadened by topographic variation through the thickness of the TEM foil. Second, based on (less pronounced) diffraction contrast parallel to the interface, the grains appear to be columnar, in many cases extending through the film thickness, with an average (lateral) grain diameter of 10--20 nm. The selected-area diffraction (SAD) pattern can be indexed according to unique (111)Py//(011)$\alpha$-W fiber texture, as shown by the hexagonal arrangement ($60^\circ$ angles) of the \{011\} reflections in $\alpha$-W, and the arrangement of \{111\} reflections in Py, $\sim 70.5^\circ$ away from the (vertical) fiber axis. The calculated diffraction spots based on \{111\}Py//\{011\}$\alpha$-W fiber texture with $\infty$-fold rotational symmetry about the film-normal axis are shown in Fig. \ref{TEM1} b), inset; good agreement is found.

\begin{figure}
  \includegraphics[width=\columnwidth]{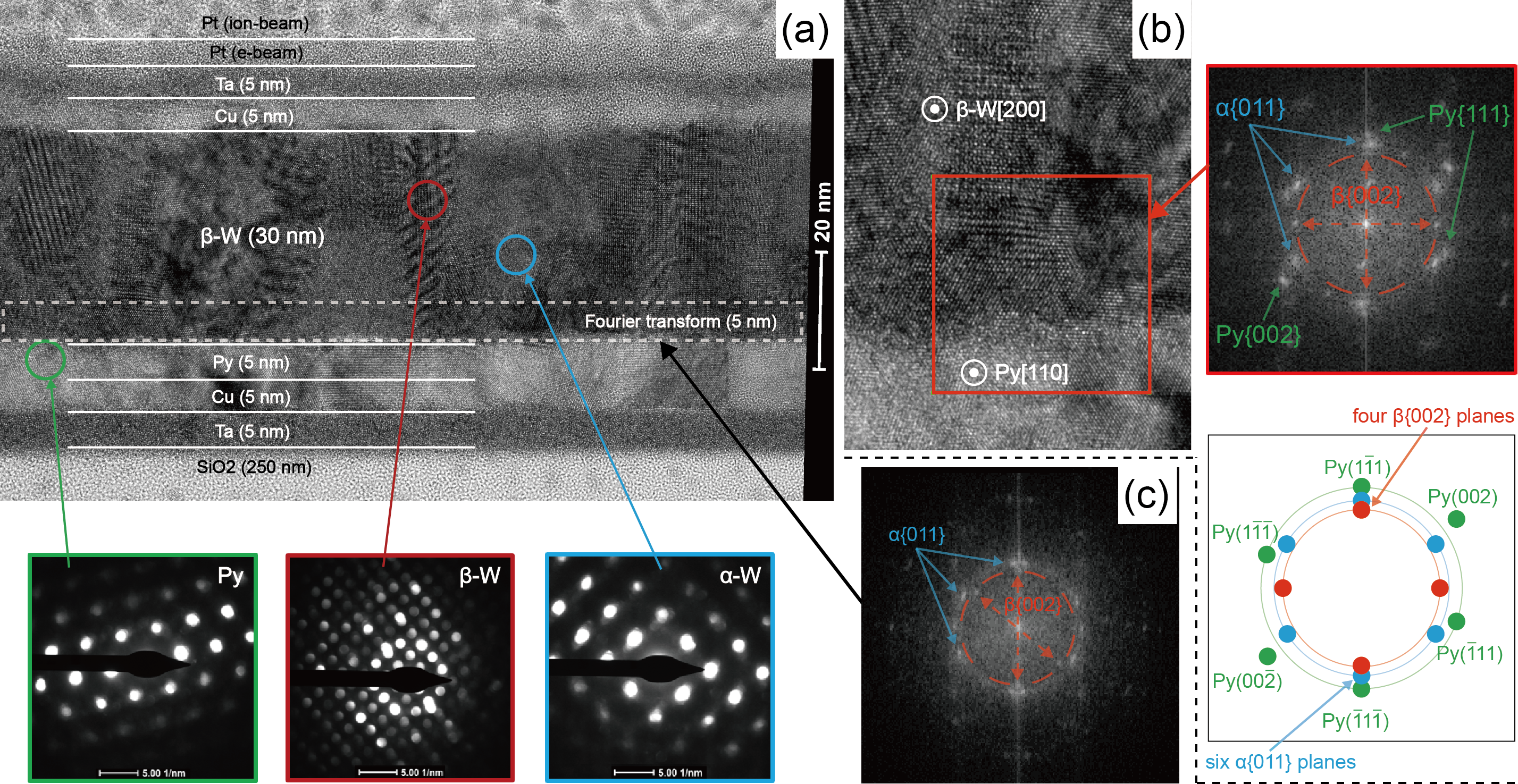}
  \caption{(a) HR-XTEM image of SiO$_{2}$/Ta(5 nm)/Cu(5 nm)/Py(5 nm)/``$\beta$''-W(30 nm)/Cu(5 nm)/Ta(5 nm), showing mixed-phase $\alpha$-W and $\beta$-W. Convergent nanobeam electron diffraction (CBED) patterns, bottom, reveal the co-existence of separated $\alpha$-W, $\beta$-W, and fcc Py. (b) Close-up of one region near the Py/W interface in (a), with discrete spatial Fourier Transform (FT). The FT is consistent with a single-crystal pattern of ($1\Bar{1}1$)[110]Py//(011)[$1\Bar{1}1$]$\alpha$-W//(002)[200]$\beta$-W, as shown in the calculated pattern (bottom right). (c) FT of interface region (dotted box), showing co-existence of $\alpha$-W and $\beta$-W in the first 5 nm W adjacent to the Py/W interface.}
  \label{TEM2}
\end{figure}

Cross-sectional images and diffraction patterns for the Py/``$\beta$''-W(30 nm) heterostructure are shown in Fig. \ref{TEM2}. Here again, in Fig. \ref{TEM2} a), the mass contrast shows similarly well-defined interfaces, but the topographic variations have a shorter wavelength, due presumably to smaller, more equiaxed grains in the mixed-phase ``$\beta$''-W. Circles indicate areas where convergent nanobeam electron diffraction (CBED) patterns were taken. The diffraction patterns over these small regions can be indexed to single phases: fcc Ni$_{81}$Fe$_{19}$ (Py) in green, bcc $\alpha$-W in blue, and A15 $\beta$-W in red.

The CBED patterns in Fig. \ref{TEM2} a) confirm that the nominal ``$\beta$''-W film is mixed-phase $\alpha$-W and $\beta$-W. The critical question for distinguishing the spin mixing conductances of $\alpha$-W and $\beta$-W in Py/W is the identity of the W phase located within the first several nanometers of the interface with Py: the pumped spin current is ejected through the interface and absorbed over this region; see the x-axis of Fig. \ref{1series}. We have addressed this question locally using high resolution imaging (see Fig. \ref{TEM2} b) and over a larger area using frequency analysis (see Fig. \ref{TEM2} c) of the image, roughly equivalent to SAD. In Fig. \ref{TEM2} b), a 10 nm area (red box) shows what appears to be a single-crystal region with ($1\Bar{1}1$)[110]Py//(011)[$1\Bar{1}1$]$\alpha$-W//(002)[200]$\beta$-W, indicating that the $\beta$ crystals may nucleate on top of the $\alpha$ crystals; however, this is contrary to our previous observations\cite{Liu_2017} and not distinguishable in the image from the superposition of grains through the foil, with nucleation of $\beta$ at the Py/W interface. The discrete spatial Fourier transform (FT) of this region shows that the four vertically/horizontally circled $\beta$-W\{002\} spots are similar in intensity to the six $\alpha$-W\{011\} spots, supporting a similar $\beta$-W content in this region. Carrying out a spatial FT of the full selected region within 5 nm of the interface (dotted box) in Fig. \ref{TEM2} a), we can confirm that $\beta$-W is indeed present adjacent to the interface, as indicated by the $\beta$-W\{002\} FT spots in Fig. \ref{TEM2} c), although these appear to be somewhat less intense than the $\alpha$-W\{011\} spots.

\subsection{Resistivity}
Four-point probe van der Pauw resistivity measurements were performed at room temperature on the first thickness series of samples ($t_{Py}=\textrm{5 nm}$ fixed, variable $t_W$) deposited on $25\times25$ mm square glass substrates, i.e., glass substrate/Ta(5 nm)/Cu(5 nm)/Py(5 nm)/W($t_W$)/Cu(5 nm)/Ta(5 nm). Two point probes for current and two point probes for voltage were placed at the four corners of the square coupons. For square samples, the voltage-to-current ratios were converted to resistance per square using the known geometrical factor $\pi / \ln{2} \approx 4.53$\cite{vanderpauw}. To isolate the W resistances, we plot the thickness-dependent sheet conductance and fit according to:

\begin{figure}
  \includegraphics[width=\columnwidth]{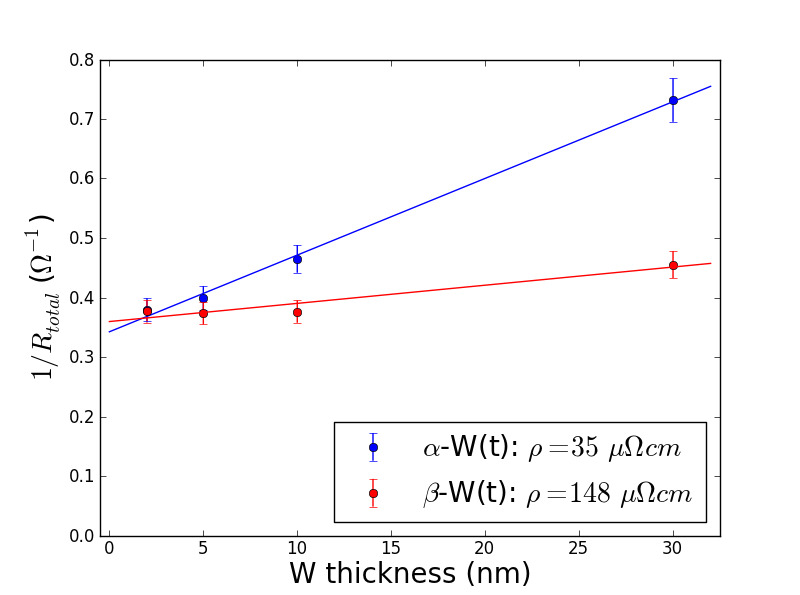}
  \caption{The total conductance $G_{total}=1/R_{total}$ as a function of W thickness. Blue dots refer to Py(5 nm)/``$\alpha$''-W($t_{W}$) samples and red dots refer to Py(5 nm)/``$\beta$''-W($t_{W}$) samples. The solid lines are linear fits.}
  \label{resistivity}
\end{figure}

\begin{equation}
\frac{1}{R_{total}}=G_{total}=G_0+\frac{4.53}{\rho_{W}}t_{W}
\label{eqn1}
\end{equation}

where $R_{total}$ ($G_{total}$) is the total resistance (conductance) of the sample, $\rho_{W}$ and $t_{W}$ are the resistivity and the thickness of the W layer, and $G_0$ is the parallel conductance of other layers in the stack.

We have verified Ohmic response by fitting the proportional dependence of voltage $V$ on current $I$ over the range 2 mA $\leq I \leq$ 10 mA for each sample. Fig. \ref{resistivity} summarizes the total conductance $G_{total}=1/R_{total}$ as a function of W thickness $t_{W}$ for all Py(5 nm)/W($t_{W}$) heterostructures. Solid lines represent linear fits for the W resistivity $\rho_W$, assumed constant as a function of W thickness for ``$\alpha$''-W and ``$\beta$''-W samples. The extracted resistivity for ``$\alpha$'' phase W $\rho_{\alpha}$ is found to be $\sim 35$ $\mu \Omega$cm and for ``$\beta$'' phase W $\rho_{\beta} \sim 148$ $\mu\Omega$cm. The resistivity for ``$\beta$''-W more than four times greater than that for ``$\alpha$''-W, is due in large part to the much smaller grain size for $\beta$-W and is typically observed in prior studies\cite{Petroff_1973}. Here the resistivity for ``$\alpha$-W'' is larger by a factor of 2--3 than films deposited at room temperature and postannealed in previous work\cite{Choi_JVST}, also attributable to a smaller grain size in these films deposited at ambient temperature. The resistivity measurements for these thin films might be taken as indirect evidence for the presence of the $\beta$ phase in the nominal ``$\beta$''-W layers.

\section{ferromagnetic resonance measurements}
The four thickness series of Py($t_{Py}$)/W($t_W$) films, for ``$\alpha$''-W and ``$\beta$''-W, as described in Section II were characterized using variable-frequency field-swept FMR using a coplanar waveguide (CPW) with center conductor width of 300 $\mu$m. The bias magnetic field was applied in the film plane ({\it pc}-FMR, or parallel condition). For details, see e.g., our prior work in Ref. [20].

\begin{figure}
  \includegraphics[width=\columnwidth]{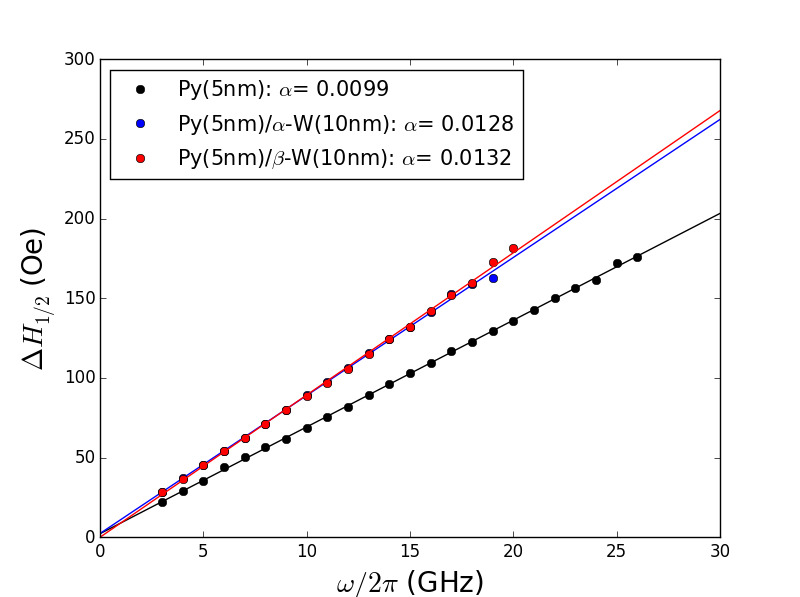}
  \caption{Half-power FMR linewidth $\Delta H_{1/2}$ spectra of reference sample Py(5 nm) (black), Py(5 nm)/``$\alpha$''-W(10 nm) (blue) and Py(5 nm)/``$\beta$''-W(10 nm) (red) samples. The solid lines are linear fits.}
  \label{linewidth}
\end{figure}

Fig. \ref{linewidth} summarizes half-power FMR linewidth $\Delta H_{1/2}$ as a function of frequency $\omega/2\pi$ for Py(5 nm), Py(5 nm)/``$\alpha$''-W(10 nm) and Py(5 nm)/``$\beta$''-W(10 nm) samples. The measurements were taken at frequencies from 3 GHz to above 20 GHz. Solid lines are linear regression of the variable-frequency FMR linewidth $\Delta H_{1/2}=\Delta H_{0}+2\alpha\omega/\gamma$, where $\Delta H_{1/2}$ is the full-width at half-maximum, $\Delta H_{0}$ is the inhomogeneous broadening, $\alpha$ is the Gilbert damping, $\omega$ is the resonance frequency and $\gamma$ is the gyromagnetic ratio. Good linear fits were obtained with resonance frequency $\omega/2\pi$ for experimental linewidths $\Delta H_{1/2}(\omega)$ of all the samples measured.

\begin{figure}
  \includegraphics[width=\columnwidth]{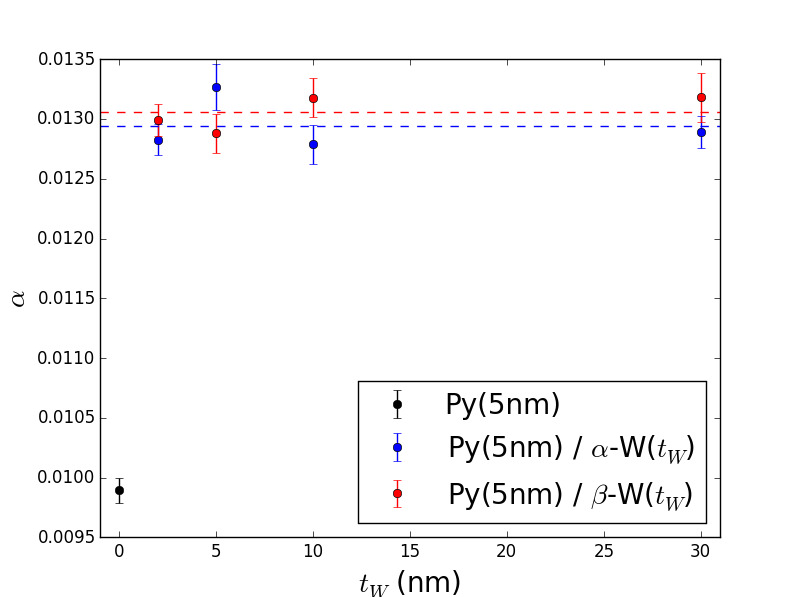}
  \caption{Gilbert damping $\alpha$ of the reference sample Py(5 nm) (black), Py(5 nm)/``$\alpha$''-W($t_{W}$) (blue) and Py(5 nm)/``$\beta$''-W($t_{W}$) (red) samples. The blue and red dash lines refer to averaged enhanced damping for Py(5 nm)/``$\alpha$''-W($t_{W}$) and Py(5 nm)/``$\beta$''-W($t_{W}$), respectively.}
  \label{1series}
\end{figure}

For the first sample thickness series Py(5 nm)/W($t_{W}$), we plot damping parameters $\alpha$ extracted from the linear fits, as a function of W thickness in Fig. \ref{1series}. Standard deviation errors in the fit for $\alpha$ are $\sim2\times10^{-4}$. The Gilbert damping $\alpha$ saturates quickly as a function of $t_{W}$ for both ``$\alpha$''-W and ``$\beta$''-W, with almost all of the effect realized with the first 2 nm of W. Loosely speaking, this fast saturation implies a short spin diffusion length $\lambda_{SD} \leq 2$ nm, so the identity of the W phase ($\alpha$ or $\beta$) over this length scale near the interface is most relevant. The averaged damping, $\alpha_{Py/``\alpha"-W}$ and $\alpha_{Py/``\beta"-W}$, are shown as horizontal dashed lines in the figure. $\alpha_{Py/``\alpha"-W}$ is slightly smaller than $\alpha_{Py/``\beta"-W}$, but this may be within experimental error. Due to spin pumping, the damping is enhanced with the addition of W layers $\Delta \alpha=\alpha_{Py/W}-\alpha_{Py}$, normalized to the Gilbert damping $\alpha_{Py}$ of the reference sample without W layers. The effective SMC $g_{eff}^{\uparrow\downarrow}$ at the Py/W interfaces can be calculated following:

\begin{equation}
\Delta \alpha=\frac{\gamma \hbar g_{eff}^{\uparrow\downarrow} }{(4\pi M_{S})t_{Py}}
\label{eqn2}
\end{equation}

where $\gamma$ is the gyromagnetic ratio, $\hbar$ is the reduced Planck constant, and $4\pi M_{S}\approx 10$ kG is the saturation inductance of Py. In this series of samples, the effective SMC at the Py/``$\alpha$''-W interface $g_{Py/``\alpha"-W}^{\uparrow\downarrow}\approx 7.2\pm0.3$ nm$^{-2}$ and the effective SMC at the Py/``$\beta$''-W interface $g_{Py/``\beta"-W}^{\uparrow\downarrow}\approx 7.4\pm0.2$ nm$^{-2}$. These values are significantly lower than those reported in Ref.\cite{Demasius_2016} for CoFeB/W (20--30 nm$^{-2}$), as measured by spin-torque FMR.

\begin{figure}
  \includegraphics[width=\columnwidth]{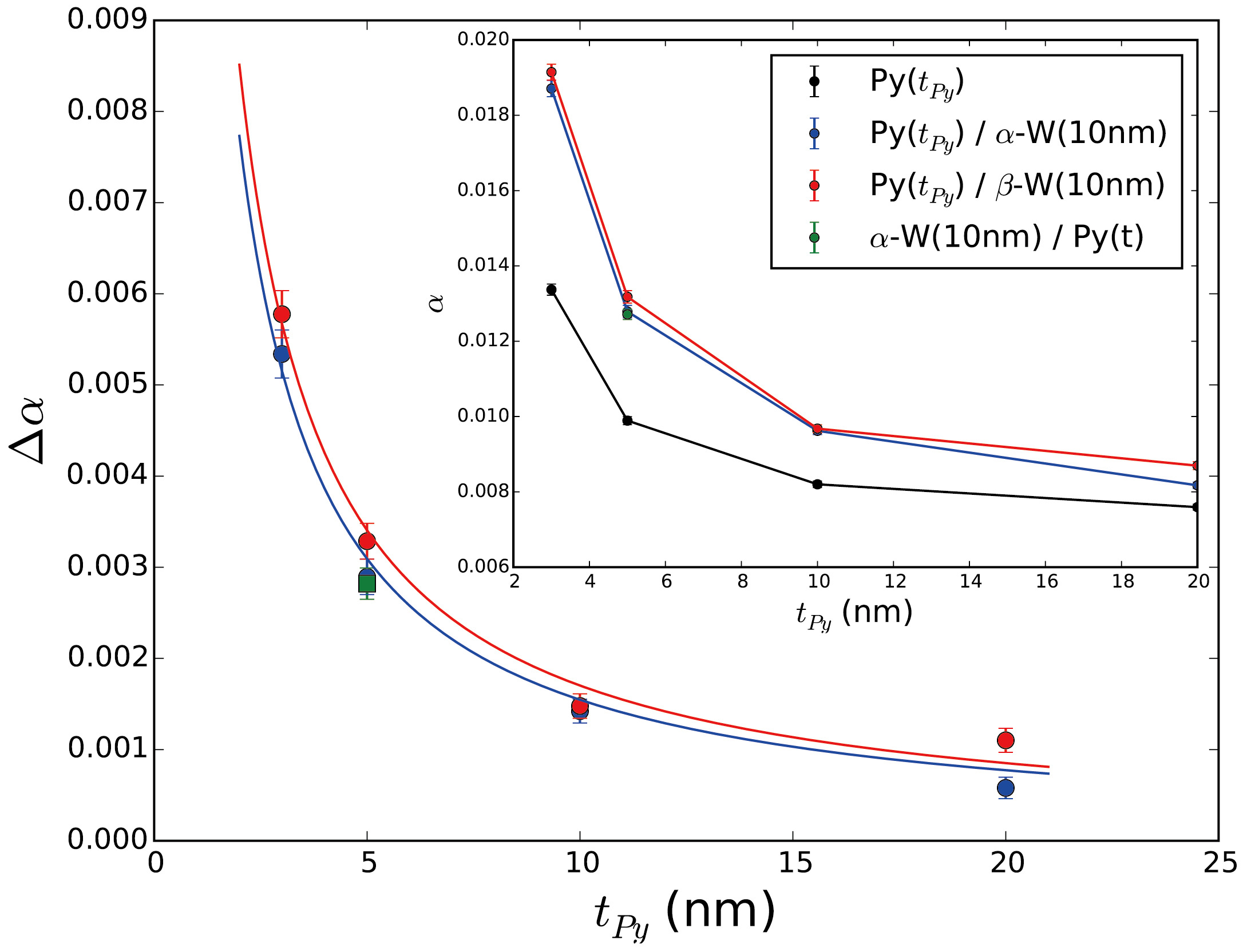}
  \caption{Damping enhancement $\Delta \alpha=\alpha_{Py/W}-\alpha_{Py}$ of Py($t_{Py}$)/``$\alpha$''-W(10 nm) (blue), Py($t_{Py}$)/``$\beta$''-W(10 nm) (red) and ``$\alpha$''-W(10 nm)/Py(5 nm) (green) samples, normalized to the Gilbert damping of reference samples $\alpha_{Py}$ with the same Py thickness. Solid lines refer to fitting with Equation \ref{eqn2}. Inset: Gilbert damping $\alpha$ of the reference sample Py($t_{Py}$) (black), Py($t_{Py}$)/``$\alpha$''-W(10 nm) (blue), Py($t_{Py}$)/``$\beta$''-W(10 nm) (red) and ``$\alpha$''-W(10 nm)/Py(5 nm) (green) samples.}
  \label{2series}
\end{figure}

For the second sample thickness series Py($t_{Py}$)/W(10 nm), we plot the extracted Gilbert damping $\alpha$ and damping enhancement $\Delta \alpha=\alpha_{Py/W}-\alpha_{Py}$ as a function of Py thickness in Fig. \ref{2series}. The enhanced damping is normalized to the Gilbert damping $\alpha_{Py}$ of reference samples with the same Py thickness $t_{Py}$. The result is in good agreement with the inverse thickness dependence of contributed damping predicted from Equation 2. The experimental data is fitted with Equation 2 to extract the effective SMC. In this series of samples, the effective SMC at the Py/``$\alpha$''-W interface $g_{Py/``\alpha"-W}^{\uparrow\downarrow}\approx 6.7\pm0.1$ nm$^{-2}$ and the effective SMC at the Py/``$\beta$''-W interface $g_{Py/``\beta"-W}^{\uparrow\downarrow}\approx 7.4\pm0.3$ nm$^{-2}$.

Previous studies on W have shown that the formation of $\alpha$-W is preferred, for thicker W layers (e.g. 10 nm)\cite{Choi_JVST,Pai_2012}. We also prepared the sample ``$\alpha$''-W(10 nm)/Py(5 nm) with reverse deposition order, with the same seed and cap layers, on an oxidized Si substrate. Here the top surface of the 10 nm thick $\alpha$-W layer is pure $\alpha$ phase, as shown by XRD in Fig. \ref{XRD} a). We performed the same FMR measurement on the reverse order sample; its Gilbert damping enhancement $\Delta \alpha$ is plotted as the green dot in Fig. \ref{2series}. This point almost overlaps with the measurement for the normal order sample Py(5 nm)/``$\alpha$''-W(10 nm), indirectly supporting the conclusion that the phase of the Py/``$\alpha$''-W interface is similar to the phase of the ``$\alpha$''-W/Py interface, i.e., almost 100\% $\alpha$ phase W. Note that it was not possible to deposit a reverse-order $\beta$ phase sample because no $\beta$ phase W could be stabilized on Cu using our technique\cite{Liu_2017}.

The FMR measurements of spin mixing conductance $g^{\uparrow\downarrow}$ for Py/``$\alpha$''-W and Py/``$\beta$''-W are new in this study. We find that the value is similar to that measured for Ta\cite{Ta2001} ($g^{\uparrow\downarrow} \sim$ 10 nm$^{-2}$) regardless of the enriched phase. First-principles-based calculations including relativistic effects\cite{Calculation2014} for $g^{\uparrow\downarrow}$ at Py/NM interfaces have shown that Ta, next to W in the periodic table, is a good spin sink due to its large spin-orbit coupling (SOC), but has a relatively small $g^{\uparrow\downarrow} \sim$ 8--9 nm$^{-2}$. The efficient absorption of spin current can be connected with a large SOC from the large atomic number, and the low SMC can be connected to relatively poor band matching across the Py/W interface, compared with that for Py/Cu or Py/Pt\cite{Calculation2014}. The conclusion for Ta is consistent with our experimental results for the Py/W system, i.e., the rapid saturation of Gilbert damping within the first 2 nm of W, indicating W is also a good spin sink, with a similarly low $g^{\uparrow\downarrow} \sim$ 7 nm$^{-2}$.

\section{discussion}
We have found very little difference between the spin scattering properties (spin mixing conductance and spin diffusion length) of $\alpha$-W and mixed phase ($\alpha$+$\beta$)-W. The simplest interpretation is that both spin mixing conductances and spin diffusion lengths are nearly equal for the two phases. However, despite our development of an optimized technique\cite{Liu_ACTA,Liu_2017,Barmak_2017} to stabilize the $\beta$ phase, our control over the amounts of deposited $\alpha$ and $\beta$ phases is less than complete, particularly near the Py/W interface.

The ``$\alpha$''-structure we deposited, Py/$\alpha$-W, is nearly $\sim$ 100\% $\alpha$ phase. We observed no strong $\beta$-W peaks in the XRD scans, and neither crystalline structure nor diffraction patterns for the $\beta$ phase in HR-XTEM characterization. According to our previous work\cite{Choi_JVST,Choi_JAP,Liu_2017}, we know that ionically and covalently bonded substrates/underlayers are favorable for the formation of some $\beta$-W, whereas metallic underlayers promote $\alpha$, so on Py even at a thickness of 2 nm, the nominally $\alpha$-W film is fully $\alpha$ if deposited in the absence of nitrogen.

In the thinnest ``$\beta$''-structure which we can characterize by XRD, Py/``$\beta$''-W(10 nm), we identify a roughly 50\%-50\% mixture of $\alpha$ and $\beta$ phases. If this balance persists at the interface as well, the SMC cannot differ by more than 10-20\% for the two phases. While the measurement of the 5 nm region near the interface seems to show somewhat less than 50\% $\beta$ phase, there is still a substantial population of $\beta$-W in this region, and it would seem that a strong difference in SMC for $\alpha$-W and $\beta$-W should be resolvable if present. Given that the measured values are very similar, we conclude that the $\alpha$ and $\beta$ phases do not differ strongly in this spin transport study.

One might ask why the spin mixing conductance, in contrast to the spin Hall angle\cite{Pai_2012}, does not differ much for the two phases of W. The spin mixing conductance (SMC) $g^{\uparrow\downarrow}_{FM/NM}$ is a property of the FM/NM interface, rather than a bulk property of the NM layer.  The SMC may be approximated (in a single-band, free-electron model) as $g^{\uparrow\downarrow}\approx \kappa k_F^2 A/4\pi^2$, where $k_F$ is the Fermi wave number for the NM, $\kappa$ represents the number of scattering channels in units of one channel per interface atom, and A is the total surface area of the interface\cite{Bauer_2002}. Despite the possibility that bulk $\beta$-W has a stronger effective spin-orbit coupling and spin Hall effect due to its A15 structure, $\beta$-W could have similar numbers of conducting channels per atom at the FM/NM interface as $\alpha$-W, which could lead to the similar values of SMC measured here.

Another possibility is that the spin diffusion length $\lambda_{SD}$ may vary along the W layer thickness, due to nonuniformly distributed $\alpha$-W and $\beta$-W phases in ``$\beta$''-W samples. If this is true, fitting a single spin diffusion length for spin pumping into very thin W layers will be problematic\cite{Montoya_PRB}. However, because we have observed a very rapid saturation of Gilbert damping over the first 2 nm of W for both ``$\alpha$''-W (almost pure $\alpha$ phase) and ``$\beta$''-W (mixed phase) in Fig. \ref{1series}, we can only assign an upper bound for $\lambda_{SD}$, similarly short in the two phases.

\section{conclusions}
In summary, we report measurements of spin mixing conductances of Py/W films with controlled amounts of $\alpha$ and $\beta$ phase W, measured by Gilbert damping through ferromagnetic resonance (FMR). We find no strong differences in the spin mixing conductances of Py/$\alpha$-W and Py/$\beta$-W, measured as $g^{\uparrow\downarrow}=$ 6.7--7.4 nm$^{-2}$, although control of the $\beta$ phase is seen to be more difficult near the interface with Py. Our experimental results also indicate that W, no matter of which phase, is a good spin sink, but with relatively small spin mixing conductance in Ni$_{81}$Fe$_{19}$ (Py)/W, similar to Ta in Py/Ta.

\section{acknowledgements}
The authors thank Daniel Paley of Columbia Nano Initiative for the grazing incidence XRD scans and Kadir Sentosun of Columbia University for the satellite peak calculations. This work is supported by the US NSF-DMR-1411160.

\theendnotes

\bibliography{ref}

\begin{thebibliography}{31}%
\makeatletter
\providecommand \@ifxundefined [1]{%
 \@ifx{#1\undefined}
}%
\providecommand \@ifnum [1]{%
 \ifnum #1\expandafter \@firstoftwo
 \else \expandafter \@secondoftwo
 \fi
}%
\providecommand \@ifx [1]{%
 \ifx #1\expandafter \@firstoftwo
 \else \expandafter \@secondoftwo
 \fi
}%
\providecommand \natexlab [1]{#1}%
\providecommand \enquote  [1]{``#1''}%
\providecommand \bibnamefont  [1]{#1}%
\providecommand \bibfnamefont [1]{#1}%
\providecommand \citenamefont [1]{#1}%
\providecommand \href@noop [0]{\@secondoftwo}%
\providecommand \href [0]{\begingroup \@sanitize@url \@href}%
\providecommand \@href[1]{\@@startlink{#1}\@@href}%
\providecommand \@@href[1]{\endgroup#1\@@endlink}%
\providecommand \@sanitize@url [0]{\catcode `\\12\catcode `\$12\catcode
  `\&12\catcode `\#12\catcode `\^12\catcode `\_12\catcode `\%12\relax}%
\providecommand \@@startlink[1]{}%
\providecommand \@@endlink[0]{}%
\providecommand \url  [0]{\begingroup\@sanitize@url \@url }%
\providecommand \@url [1]{\endgroup\@href {#1}{\urlprefix }}%
\providecommand \urlprefix  [0]{URL }%
\providecommand \Eprint [0]{\href }%
\providecommand \doibase [0]{http://dx.doi.org/}%
\providecommand \selectlanguage [0]{\@gobble}%
\providecommand \bibinfo  [0]{\@secondoftwo}%
\providecommand \bibfield  [0]{\@secondoftwo}%
\providecommand \translation [1]{[#1]}%
\providecommand \BibitemOpen [0]{}%
\providecommand \bibitemStop [0]{}%
\providecommand \bibitemNoStop [0]{.\EOS\space}%
\providecommand \EOS [0]{\spacefactor3000\relax}%
\providecommand \BibitemShut  [1]{\csname bibitem#1\endcsname}%
\let\auto@bib@innerbib\@empty
\bibitem [{\citenamefont {Saitoh}\ \emph {et~al.}(2006)\citenamefont {Saitoh},
  \citenamefont {Ueda}, \citenamefont {Miyajima},\ and\ \citenamefont
  {Tatara}}]{Saitoh_2006}%
  \BibitemOpen
  \bibfield  {author} {\bibinfo {author} {\bibfnamefont {E.}~\bibnamefont
  {Saitoh}}, \bibinfo {author} {\bibfnamefont {M.}~\bibnamefont {Ueda}},
  \bibinfo {author} {\bibfnamefont {H.}~\bibnamefont {Miyajima}}, \ and\
  \bibinfo {author} {\bibfnamefont {G.}~\bibnamefont {Tatara}},\ }\href
  {\doibase 10.1063/1.2199473} {\bibfield  {journal} {\bibinfo  {journal}
  {Applied Physics Letters}\ }\textbf {\bibinfo {volume} {88}},\ \bibinfo
  {pages} {182509} (\bibinfo {year} {2006})}\BibitemShut {NoStop}%
\bibitem [{\citenamefont {Liu}\ \emph {et~al.}(2011)\citenamefont {Liu},
  \citenamefont {Moriyama}, \citenamefont {Ralph},\ and\ \citenamefont
  {Buhrman}}]{Liu_2011}%
  \BibitemOpen
  \bibfield  {author} {\bibinfo {author} {\bibfnamefont {L.}~\bibnamefont
  {Liu}}, \bibinfo {author} {\bibfnamefont {T.}~\bibnamefont {Moriyama}},
  \bibinfo {author} {\bibfnamefont {D.~C.}\ \bibnamefont {Ralph}}, \ and\
  \bibinfo {author} {\bibfnamefont {R.~A.}\ \bibnamefont {Buhrman}},\ }\href
  {\doibase 10.1103/PhysRevLett.106.036601} {\bibfield  {journal} {\bibinfo
  {journal} {Phys. Rev. Lett.}\ }\textbf {\bibinfo {volume} {106}},\ \bibinfo
  {pages} {036601} (\bibinfo {year} {2011})}\BibitemShut {NoStop}%
\bibitem [{\citenamefont {Pai}\ \emph {et~al.}(2012)\citenamefont {Pai},
  \citenamefont {Liu}, \citenamefont {Li}, \citenamefont {Tseng}, \citenamefont
  {Ralph},\ and\ \citenamefont {Buhrman}}]{Pai_2012}%
  \BibitemOpen
  \bibfield  {author} {\bibinfo {author} {\bibfnamefont {C.-F.}\ \bibnamefont
  {Pai}}, \bibinfo {author} {\bibfnamefont {L.}~\bibnamefont {Liu}}, \bibinfo
  {author} {\bibfnamefont {Y.}~\bibnamefont {Li}}, \bibinfo {author}
  {\bibfnamefont {H.~W.}\ \bibnamefont {Tseng}}, \bibinfo {author}
  {\bibfnamefont {D.~C.}\ \bibnamefont {Ralph}}, \ and\ \bibinfo {author}
  {\bibfnamefont {R.~A.}\ \bibnamefont {Buhrman}},\ }\href {\doibase
  http://dx.doi.org/10.1063/1.4753947} {\bibfield  {journal} {\bibinfo
  {journal} {Applied Physics Letters}\ }\textbf {\bibinfo {volume} {101}},\
  \bibinfo {pages} {122404} (\bibinfo {year} {2012})}\BibitemShut {NoStop}%
\bibitem [{\citenamefont {Wang}\ \emph {et~al.}(2014)\citenamefont {Wang},
  \citenamefont {Du}, \citenamefont {Pu}, \citenamefont {Adur}, \citenamefont
  {Hammel},\ and\ \citenamefont {Yang}}]{Wang_2014}%
  \BibitemOpen
  \bibfield  {author} {\bibinfo {author} {\bibfnamefont {H.~L.}\ \bibnamefont
  {Wang}}, \bibinfo {author} {\bibfnamefont {C.~H.}\ \bibnamefont {Du}},
  \bibinfo {author} {\bibfnamefont {Y.}~\bibnamefont {Pu}}, \bibinfo {author}
  {\bibfnamefont {R.}~\bibnamefont {Adur}}, \bibinfo {author} {\bibfnamefont
  {P.~C.}\ \bibnamefont {Hammel}}, \ and\ \bibinfo {author} {\bibfnamefont
  {F.~Y.}\ \bibnamefont {Yang}},\ }\href {\doibase
  10.1103/PhysRevLett.112.197201} {\bibfield  {journal} {\bibinfo  {journal}
  {Phys. Rev. Lett.}\ }\textbf {\bibinfo {volume} {112}},\ \bibinfo {pages}
  {197201} (\bibinfo {year} {2014})}\BibitemShut {NoStop}%
\bibitem [{\citenamefont {Hartmann}, \citenamefont {Ebert},\ and\ \citenamefont
  {Bretschneider}(1931)}]{A15bW}%
  \BibitemOpen
  \bibfield  {author} {\bibinfo {author} {\bibfnamefont {H.}~\bibnamefont
  {Hartmann}}, \bibinfo {author} {\bibfnamefont {F.}~\bibnamefont {Ebert}}, \
  and\ \bibinfo {author} {\bibfnamefont {O.}~\bibnamefont {Bretschneider}},\
  }\href {\doibase 10.1002/zaac.19311980111} {\bibfield  {journal} {\bibinfo
  {journal} {Z. Anorg. Allg. Chem.}\ }\textbf {\bibinfo {volume} {198}},\
  \bibinfo {pages} {116} (\bibinfo {year} {1931})}\BibitemShut {NoStop}%
\bibitem [{\citenamefont {Hao}\ and\ \citenamefont {Xiao}(2015)}]{Hao_2015}%
  \BibitemOpen
  \bibfield  {author} {\bibinfo {author} {\bibfnamefont {Q.}~\bibnamefont
  {Hao}}\ and\ \bibinfo {author} {\bibfnamefont {G.}~\bibnamefont {Xiao}},\
  }\href {\doibase 10.1103/PhysRevApplied.3.034009} {\bibfield  {journal}
  {\bibinfo  {journal} {Phys. Rev. Applied}\ }\textbf {\bibinfo {volume} {3}},\
  \bibinfo {pages} {034009} (\bibinfo {year} {2015})}\BibitemShut {NoStop}%
\bibitem [{\citenamefont {Liu}\ \emph {et~al.}(2015)\citenamefont {Liu},
  \citenamefont {Ohkubo}, \citenamefont {Mitani}, \citenamefont {Hono},\ and\
  \citenamefont {Hayashi}}]{Liu_2015}%
  \BibitemOpen
  \bibfield  {author} {\bibinfo {author} {\bibfnamefont {J.}~\bibnamefont
  {Liu}}, \bibinfo {author} {\bibfnamefont {T.}~\bibnamefont {Ohkubo}},
  \bibinfo {author} {\bibfnamefont {S.}~\bibnamefont {Mitani}}, \bibinfo
  {author} {\bibfnamefont {K.}~\bibnamefont {Hono}}, \ and\ \bibinfo {author}
  {\bibfnamefont {M.}~\bibnamefont {Hayashi}},\ }\href {\doibase
  10.1063/1.4937452} {\bibfield  {journal} {\bibinfo  {journal} {Applied
  Physics Letters}\ }\textbf {\bibinfo {volume} {107}},\ \bibinfo {pages}
  {232408} (\bibinfo {year} {2015})}\BibitemShut {NoStop}%
\bibitem [{\citenamefont {Demasius}\ \emph {et~al.}(2016)\citenamefont
  {Demasius}, \citenamefont {Phung}, \citenamefont {Zhang}, \citenamefont
  {Hughes}, \citenamefont {Yang}, \citenamefont {Kellock}, \citenamefont {Han},
  \citenamefont {Pushp},\ and\ \citenamefont {Parkin}}]{Demasius_2016}%
  \BibitemOpen
  \bibfield  {author} {\bibinfo {author} {\bibfnamefont {K.-U.}\ \bibnamefont
  {Demasius}}, \bibinfo {author} {\bibfnamefont {T.}~\bibnamefont {Phung}},
  \bibinfo {author} {\bibfnamefont {W.}~\bibnamefont {Zhang}}, \bibinfo
  {author} {\bibfnamefont {B.~P.}\ \bibnamefont {Hughes}}, \bibinfo {author}
  {\bibfnamefont {S.-H.}\ \bibnamefont {Yang}}, \bibinfo {author}
  {\bibfnamefont {A.}~\bibnamefont {Kellock}}, \bibinfo {author} {\bibfnamefont
  {W.}~\bibnamefont {Han}}, \bibinfo {author} {\bibfnamefont {A.}~\bibnamefont
  {Pushp}}, \ and\ \bibinfo {author} {\bibfnamefont {S.~S.~P.}\ \bibnamefont
  {Parkin}},\ }\href {\doibase 10.1038/ncomms10644} {\bibfield  {journal}
  {\bibinfo  {journal} {Nature Communications}\ }\textbf {\bibinfo {volume}
  {7}},\ \bibinfo {pages} {10644} (\bibinfo {year} {2016})}\BibitemShut
  {NoStop}%
\bibitem [{\citenamefont {Liu}\ and\ \citenamefont {Barmak}(2016)}]{Liu_ACTA}%
  \BibitemOpen
  \bibfield  {author} {\bibinfo {author} {\bibfnamefont {J.}~\bibnamefont
  {Liu}}\ and\ \bibinfo {author} {\bibfnamefont {K.}~\bibnamefont {Barmak}},\
  }\href {\doibase https://doi.org/10.1016/j.actamat.2015.11.049} {\bibfield
  {journal} {\bibinfo  {journal} {Acta Materialia}\ }\textbf {\bibinfo {volume}
  {104}},\ \bibinfo {pages} {223 } (\bibinfo {year} {2016})}\BibitemShut
  {NoStop}%
\bibitem [{\citenamefont {Barmak}\ and\ \citenamefont {Liu}(2017)}]{Liu_2017}%
  \BibitemOpen
  \bibfield  {author} {\bibinfo {author} {\bibfnamefont {K.}~\bibnamefont
  {Barmak}}\ and\ \bibinfo {author} {\bibfnamefont {J.}~\bibnamefont {Liu}},\
  }\href {\doibase 10.1116/1.5003628} {\bibfield  {journal} {\bibinfo
  {journal} {Journal of Vacuum Science {\&} Technology A: Vacuum, Surfaces, and
  Films}\ }\textbf {\bibinfo {volume} {35}},\ \bibinfo {pages} {061516}
  (\bibinfo {year} {2017})}\BibitemShut {NoStop}%
\bibitem [{\citenamefont {Barmak}\ \emph {et~al.}(2017)\citenamefont {Barmak},
  \citenamefont {Liu}, \citenamefont {Harlan}, \citenamefont {Xiao},
  \citenamefont {Duncan},\ and\ \citenamefont {Henkelman}}]{Barmak_2017}%
  \BibitemOpen
  \bibfield  {author} {\bibinfo {author} {\bibfnamefont {K.}~\bibnamefont
  {Barmak}}, \bibinfo {author} {\bibfnamefont {J.}~\bibnamefont {Liu}},
  \bibinfo {author} {\bibfnamefont {L.}~\bibnamefont {Harlan}}, \bibinfo
  {author} {\bibfnamefont {P.}~\bibnamefont {Xiao}}, \bibinfo {author}
  {\bibfnamefont {J.}~\bibnamefont {Duncan}}, \ and\ \bibinfo {author}
  {\bibfnamefont {G.}~\bibnamefont {Henkelman}},\ }\href {\doibase
  https://doi.org/10.1063/1.4995261} {\bibfield  {journal} {\bibinfo  {journal}
  {The Journal of Chemical Physics}\ }\textbf {\bibinfo {volume} {147}},\
  \bibinfo {pages} {152709} (\bibinfo {year} {2017})}\BibitemShut {NoStop}%
\bibitem [{\citenamefont {Arita}\ and\ \citenamefont
  {Nishida}(1993)}]{Arita_1993}%
  \BibitemOpen
  \bibfield  {author} {\bibinfo {author} {\bibfnamefont {M.}~\bibnamefont
  {Arita}}\ and\ \bibinfo {author} {\bibfnamefont {I.}~\bibnamefont
  {Nishida}},\ }\href {\doibase 10.1143/jjap.32.1759} {\bibfield  {journal}
  {\bibinfo  {journal} {Japanese Journal of Applied Physics}\ }\textbf
  {\bibinfo {volume} {32}},\ \bibinfo {pages} {1759} (\bibinfo {year}
  {1993})}\BibitemShut {NoStop}%
\bibitem [{\citenamefont {H{\"{a}}gg}\ and\ \citenamefont
  {Sch{\"{o}}nberg}(1954)}]{Hagg1954}%
  \BibitemOpen
  \bibfield  {author} {\bibinfo {author} {\bibfnamefont {G.}~\bibnamefont
  {H{\"{a}}gg}}\ and\ \bibinfo {author} {\bibfnamefont {N.}~\bibnamefont
  {Sch{\"{o}}nberg}},\ }\href {\doibase 10.1107/S0365110X54000989} {\bibfield
  {journal} {\bibinfo  {journal} {Acta Crystallographica}\ }\textbf {\bibinfo
  {volume} {7}},\ \bibinfo {pages} {351} (\bibinfo {year} {1954})}\BibitemShut
  {NoStop}%
\bibitem [{\citenamefont {Urban}, \citenamefont {Woltersdorf},\ and\
  \citenamefont {Heinrich}(2001)}]{Heinrich2001}%
  \BibitemOpen
  \bibfield  {author} {\bibinfo {author} {\bibfnamefont {R.}~\bibnamefont
  {Urban}}, \bibinfo {author} {\bibfnamefont {G.}~\bibnamefont {Woltersdorf}},
  \ and\ \bibinfo {author} {\bibfnamefont {B.}~\bibnamefont {Heinrich}},\
  }\href {\doibase 10.1103/PhysRevLett.87.217204} {\bibfield  {journal}
  {\bibinfo  {journal} {Phys. Rev. Lett.}\ }\textbf {\bibinfo {volume} {87}},\
  \bibinfo {pages} {217204} (\bibinfo {year} {2001})}\BibitemShut {NoStop}%
\bibitem [{\citenamefont {Tserkovnyak}, \citenamefont {Brataas},\ and\
  \citenamefont {Bauer}(2002{\natexlab{a}})}]{Tserkovnyak2002}%
  \BibitemOpen
  \bibfield  {author} {\bibinfo {author} {\bibfnamefont {Y.}~\bibnamefont
  {Tserkovnyak}}, \bibinfo {author} {\bibfnamefont {A.}~\bibnamefont
  {Brataas}}, \ and\ \bibinfo {author} {\bibfnamefont {G.~E.~W.}\ \bibnamefont
  {Bauer}},\ }\href {\doibase 10.1103/PhysRevLett.88.117601} {\bibfield
  {journal} {\bibinfo  {journal} {Phys. Rev. Lett.}\ }\textbf {\bibinfo
  {volume} {88}},\ \bibinfo {pages} {117601} (\bibinfo {year}
  {2002}{\natexlab{a}})}\BibitemShut {NoStop}%
\bibitem [{\citenamefont {Mosendz}\ \emph {et~al.}(2010)\citenamefont
  {Mosendz}, \citenamefont {Pearson}, \citenamefont {Fradin}, \citenamefont
  {Bauer}, \citenamefont {Bader},\ and\ \citenamefont
  {Hoffmann}}]{Mosendz2010}%
  \BibitemOpen
  \bibfield  {author} {\bibinfo {author} {\bibfnamefont {O.}~\bibnamefont
  {Mosendz}}, \bibinfo {author} {\bibfnamefont {J.~E.}\ \bibnamefont
  {Pearson}}, \bibinfo {author} {\bibfnamefont {F.~Y.}\ \bibnamefont {Fradin}},
  \bibinfo {author} {\bibfnamefont {G.~E.~W.}\ \bibnamefont {Bauer}}, \bibinfo
  {author} {\bibfnamefont {S.~D.}\ \bibnamefont {Bader}}, \ and\ \bibinfo
  {author} {\bibfnamefont {A.}~\bibnamefont {Hoffmann}},\ }\href {\doibase
  10.1103/PhysRevLett.104.046601} {\bibfield  {journal} {\bibinfo  {journal}
  {Phys. Rev. Lett.}\ }\textbf {\bibinfo {volume} {104}},\ \bibinfo {pages}
  {046601} (\bibinfo {year} {2010})}\BibitemShut {NoStop}%
\bibitem [{\citenamefont {Caminale}\ \emph {et~al.}(2016)\citenamefont
  {Caminale}, \citenamefont {Ghosh}, \citenamefont {Auffret}, \citenamefont
  {Ebels}, \citenamefont {Ollefs}, \citenamefont {Wilhelm}, \citenamefont
  {Rogalev},\ and\ \citenamefont {Bailey}}]{Caminale2016}%
  \BibitemOpen
  \bibfield  {author} {\bibinfo {author} {\bibfnamefont {M.}~\bibnamefont
  {Caminale}}, \bibinfo {author} {\bibfnamefont {A.}~\bibnamefont {Ghosh}},
  \bibinfo {author} {\bibfnamefont {S.}~\bibnamefont {Auffret}}, \bibinfo
  {author} {\bibfnamefont {U.}~\bibnamefont {Ebels}}, \bibinfo {author}
  {\bibfnamefont {K.}~\bibnamefont {Ollefs}}, \bibinfo {author} {\bibfnamefont
  {F.}~\bibnamefont {Wilhelm}}, \bibinfo {author} {\bibfnamefont
  {A.}~\bibnamefont {Rogalev}}, \ and\ \bibinfo {author} {\bibfnamefont
  {W.~E.}\ \bibnamefont {Bailey}},\ }\href {\doibase
  10.1103/PhysRevB.94.014414} {\bibfield  {journal} {\bibinfo  {journal} {Phys.
  Rev. B}\ }\textbf {\bibinfo {volume} {94}},\ \bibinfo {pages} {014414}
  (\bibinfo {year} {2016})}\BibitemShut {NoStop}%
\bibitem [{\citenamefont {Li}, \citenamefont {Cao},\ and\ \citenamefont
  {Bailey}(2016)}]{Yi2016}%
  \BibitemOpen
  \bibfield  {author} {\bibinfo {author} {\bibfnamefont {Y.}~\bibnamefont
  {Li}}, \bibinfo {author} {\bibfnamefont {W.}~\bibnamefont {Cao}}, \ and\
  \bibinfo {author} {\bibfnamefont {W.~E.}\ \bibnamefont {Bailey}},\ }\href
  {\doibase 10.1103/PhysRevB.94.174439} {\bibfield  {journal} {\bibinfo
  {journal} {Phys. Rev. B}\ }\textbf {\bibinfo {volume} {94}},\ \bibinfo
  {pages} {174439} (\bibinfo {year} {2016})}\BibitemShut {NoStop}%
\bibitem [{\citenamefont {Cao}\ \emph {et~al.}(2019)\citenamefont {Cao},
  \citenamefont {Yang}, \citenamefont {Auffret},\ and\ \citenamefont
  {Bailey}}]{Wei2019}%
  \BibitemOpen
  \bibfield  {author} {\bibinfo {author} {\bibfnamefont {W.}~\bibnamefont
  {Cao}}, \bibinfo {author} {\bibfnamefont {L.}~\bibnamefont {Yang}}, \bibinfo
  {author} {\bibfnamefont {S.}~\bibnamefont {Auffret}}, \ and\ \bibinfo
  {author} {\bibfnamefont {W.~E.}\ \bibnamefont {Bailey}},\ }\href {\doibase
  10.1103/PhysRevB.99.094406} {\bibfield  {journal} {\bibinfo  {journal} {Phys.
  Rev. B}\ }\textbf {\bibinfo {volume} {99}},\ \bibinfo {pages} {094406}
  (\bibinfo {year} {2019})}\BibitemShut {NoStop}%
\bibitem [{\citenamefont {Otte}(1961)}]{Otte1961}%
  \BibitemOpen
  \bibfield  {author} {\bibinfo {author} {\bibfnamefont {H.~M.}\ \bibnamefont
  {Otte}},\ }\href {\doibase 10.1063/1.1728392} {\bibfield  {journal} {\bibinfo
   {journal} {Journal of Applied Physics}\ }\textbf {\bibinfo {volume} {32}},\
  \bibinfo {pages} {1536} (\bibinfo {year} {1961})}\BibitemShut {NoStop}%
\bibitem [{\citenamefont {Huang}\ \emph {et~al.}(1997)\citenamefont {Huang},
  \citenamefont {Wang}, \citenamefont {Yu}, \citenamefont {Hu}, \citenamefont
  {Lee},\ and\ \citenamefont {Yang}}]{HUANG_1997}%
  \BibitemOpen
  \bibfield  {author} {\bibinfo {author} {\bibfnamefont {J.}~\bibnamefont
  {Huang}}, \bibinfo {author} {\bibfnamefont {T.}~\bibnamefont {Wang}},
  \bibinfo {author} {\bibfnamefont {C.}~\bibnamefont {Yu}}, \bibinfo {author}
  {\bibfnamefont {Y.}~\bibnamefont {Hu}}, \bibinfo {author} {\bibfnamefont
  {P.}~\bibnamefont {Lee}}, \ and\ \bibinfo {author} {\bibfnamefont
  {M.}~\bibnamefont {Yang}},\ }\href {\doibase
  https://doi.org/10.1016/S0022-0248(96)00694-X} {\bibfield  {journal}
  {\bibinfo  {journal} {Journal of Crystal Growth}\ }\textbf {\bibinfo {volume}
  {171}},\ \bibinfo {pages} {442 } (\bibinfo {year} {1997})}\BibitemShut
  {NoStop}%
\bibitem [{\citenamefont {Scherrer}(1918)}]{scherrer1918}%
  \BibitemOpen
  \bibfield  {author} {\bibinfo {author} {\bibfnamefont {P.}~\bibnamefont
  {Scherrer}},\ }\href@noop {} {\bibfield  {journal} {\bibinfo  {journal}
  {G{\"o}ttinger Nachrichten Math. Phys.}\ }\textbf {\bibinfo {volume} {2}},\
  \bibinfo {pages} {98} (\bibinfo {year} {1918})}\BibitemShut {NoStop}%
\bibitem [{\citenamefont {Ying}, \citenamefont {Murray},\ and\ \citenamefont
  {Noyan}(2009)}]{Ying_2009}%
  \BibitemOpen
  \bibfield  {author} {\bibinfo {author} {\bibfnamefont {A.~J.}\ \bibnamefont
  {Ying}}, \bibinfo {author} {\bibfnamefont {C.~E.}\ \bibnamefont {Murray}}, \
  and\ \bibinfo {author} {\bibfnamefont {I.~C.}\ \bibnamefont {Noyan}},\ }\href
  {\doibase 10.1107/S0021889809006888} {\bibfield  {journal} {\bibinfo
  {journal} {Journal of Applied Crystallography}\ }\textbf {\bibinfo {volume}
  {42}},\ \bibinfo {pages} {401} (\bibinfo {year} {2009})}\BibitemShut
  {NoStop}%
\bibitem [{\citenamefont {van~der Pauw}(1958)}]{vanderpauw}%
  \BibitemOpen
  \bibfield  {author} {\bibinfo {author} {\bibfnamefont {L.~J.}\ \bibnamefont
  {van~der Pauw}},\ }\href {https://ci.nii.ac.jp/naid/20000484641/en/}
  {\bibfield  {journal} {\bibinfo  {journal} {Philips Tech. Rev.}\ }\textbf
  {\bibinfo {volume} {20}},\ \bibinfo {pages} {220} (\bibinfo {year}
  {1958})}\BibitemShut {NoStop}%
\bibitem [{\citenamefont {Petroff}\ \emph {et~al.}(1973)\citenamefont
  {Petroff}, \citenamefont {Sheng}, \citenamefont {Sinha}, \citenamefont
  {Rozgonyi},\ and\ \citenamefont {Alexander}}]{Petroff_1973}%
  \BibitemOpen
  \bibfield  {author} {\bibinfo {author} {\bibfnamefont {P.}~\bibnamefont
  {Petroff}}, \bibinfo {author} {\bibfnamefont {T.~T.}\ \bibnamefont {Sheng}},
  \bibinfo {author} {\bibfnamefont {A.~K.}\ \bibnamefont {Sinha}}, \bibinfo
  {author} {\bibfnamefont {G.~A.}\ \bibnamefont {Rozgonyi}}, \ and\ \bibinfo
  {author} {\bibfnamefont {F.~B.}\ \bibnamefont {Alexander}},\ }\href {\doibase
  https://doi.org/10.1063/1.1662611} {\bibfield  {journal} {\bibinfo  {journal}
  {Journal of Applied Physics}\ }\textbf {\bibinfo {volume} {44}},\ \bibinfo
  {pages} {2545} (\bibinfo {year} {1973})}\BibitemShut {NoStop}%
\bibitem [{\citenamefont {Choi}\ \emph {et~al.}(2011)\citenamefont {Choi},
  \citenamefont {Wang}, \citenamefont {Chung}, \citenamefont {Liu},
  \citenamefont {Darbal}, \citenamefont {Wise}, \citenamefont {Nuhfer},
  \citenamefont {Barmak}, \citenamefont {Warren}, \citenamefont {Coffey},\ and\
  \citenamefont {Toney}}]{Choi_JVST}%
  \BibitemOpen
  \bibfield  {author} {\bibinfo {author} {\bibfnamefont {D.}~\bibnamefont
  {Choi}}, \bibinfo {author} {\bibfnamefont {B.}~\bibnamefont {Wang}}, \bibinfo
  {author} {\bibfnamefont {S.}~\bibnamefont {Chung}}, \bibinfo {author}
  {\bibfnamefont {X.}~\bibnamefont {Liu}}, \bibinfo {author} {\bibfnamefont
  {A.}~\bibnamefont {Darbal}}, \bibinfo {author} {\bibfnamefont
  {A.}~\bibnamefont {Wise}}, \bibinfo {author} {\bibfnamefont {N.~T.}\
  \bibnamefont {Nuhfer}}, \bibinfo {author} {\bibfnamefont {K.}~\bibnamefont
  {Barmak}}, \bibinfo {author} {\bibfnamefont {A.~P.}\ \bibnamefont {Warren}},
  \bibinfo {author} {\bibfnamefont {K.~R.}\ \bibnamefont {Coffey}}, \ and\
  \bibinfo {author} {\bibfnamefont {M.~F.}\ \bibnamefont {Toney}},\ }\href
  {\doibase 10.1116/1.3622619} {\bibfield  {journal} {\bibinfo  {journal}
  {Journal of Vacuum Science {\&} Technology A: Vacuum, Surfaces, and Films}\
  }\textbf {\bibinfo {volume} {29}},\ \bibinfo {pages} {051512} (\bibinfo
  {year} {2011})}\BibitemShut {NoStop}%
\bibitem [{\citenamefont {Mizukami}, \citenamefont {Ando},\ and\ \citenamefont
  {Miyazaki}(2001)}]{Ta2001}%
  \BibitemOpen
  \bibfield  {author} {\bibinfo {author} {\bibfnamefont {S.}~\bibnamefont
  {Mizukami}}, \bibinfo {author} {\bibfnamefont {Y.}~\bibnamefont {Ando}}, \
  and\ \bibinfo {author} {\bibfnamefont {T.}~\bibnamefont {Miyazaki}},\ }\href
  {\doibase https://doi.org/10.1016/S0304-8853(00)01097-0} {\bibfield
  {journal} {\bibinfo  {journal} {Journal of Magnetism and Magnetic Materials}\
  }\textbf {\bibinfo {volume} {226-230}},\ \bibinfo {pages} {1640 } (\bibinfo
  {year} {2001})},\ \bibinfo {note} {proceedings of the International
  Conference on Magnetism (ICM 2000)}\BibitemShut {NoStop}%
\bibitem [{\citenamefont {Liu}\ \emph {et~al.}(2014)\citenamefont {Liu},
  \citenamefont {Yuan}, \citenamefont {Wesselink}, \citenamefont {Starikov},\
  and\ \citenamefont {Kelly}}]{Calculation2014}%
  \BibitemOpen
  \bibfield  {author} {\bibinfo {author} {\bibfnamefont {Y.}~\bibnamefont
  {Liu}}, \bibinfo {author} {\bibfnamefont {Z.}~\bibnamefont {Yuan}}, \bibinfo
  {author} {\bibfnamefont {R.~J.~H.}\ \bibnamefont {Wesselink}}, \bibinfo
  {author} {\bibfnamefont {A.~A.}\ \bibnamefont {Starikov}}, \ and\ \bibinfo
  {author} {\bibfnamefont {P.~J.}\ \bibnamefont {Kelly}},\ }\href {\doibase
  10.1103/PhysRevLett.113.207202} {\bibfield  {journal} {\bibinfo  {journal}
  {Phys. Rev. Lett.}\ }\textbf {\bibinfo {volume} {113}},\ \bibinfo {pages}
  {207202} (\bibinfo {year} {2014})}\BibitemShut {NoStop}%
\bibitem [{\citenamefont {Choi}\ \emph {et~al.}(2014)\citenamefont {Choi},
  \citenamefont {Liu}, \citenamefont {Schelling}, \citenamefont {Coffey},\ and\
  \citenamefont {Barmak}}]{Choi_JAP}%
  \BibitemOpen
  \bibfield  {author} {\bibinfo {author} {\bibfnamefont {D.}~\bibnamefont
  {Choi}}, \bibinfo {author} {\bibfnamefont {X.}~\bibnamefont {Liu}}, \bibinfo
  {author} {\bibfnamefont {P.~K.}\ \bibnamefont {Schelling}}, \bibinfo {author}
  {\bibfnamefont {K.~R.}\ \bibnamefont {Coffey}}, \ and\ \bibinfo {author}
  {\bibfnamefont {K.}~\bibnamefont {Barmak}},\ }\href {\doibase
  10.1063/1.4868093} {\bibfield  {journal} {\bibinfo  {journal} {Journal of
  Applied Physics}\ }\textbf {\bibinfo {volume} {115}},\ \bibinfo {pages}
  {104308} (\bibinfo {year} {2014})}\BibitemShut {NoStop}%
\bibitem [{\citenamefont {Tserkovnyak}, \citenamefont {Brataas},\ and\
  \citenamefont {Bauer}(2002{\natexlab{b}})}]{Bauer_2002}%
  \BibitemOpen
  \bibfield  {author} {\bibinfo {author} {\bibfnamefont {Y.}~\bibnamefont
  {Tserkovnyak}}, \bibinfo {author} {\bibfnamefont {A.}~\bibnamefont
  {Brataas}}, \ and\ \bibinfo {author} {\bibfnamefont {G.~E.~W.}\ \bibnamefont
  {Bauer}},\ }\href {\doibase 10.1103/PhysRevLett.88.117601} {\bibfield
  {journal} {\bibinfo  {journal} {Phys. Rev. Lett.}\ }\textbf {\bibinfo
  {volume} {88}},\ \bibinfo {pages} {117601} (\bibinfo {year}
  {2002}{\natexlab{b}})}\BibitemShut {NoStop}%
\bibitem [{\citenamefont {Montoya}\ \emph {et~al.}(2016)\citenamefont
  {Montoya}, \citenamefont {Omelchenko}, \citenamefont {Coutts}, \citenamefont
  {Lee-Hone}, \citenamefont {H\"ubner}, \citenamefont {Broun}, \citenamefont
  {Heinrich},\ and\ \citenamefont {Girt}}]{Montoya_PRB}%
  \BibitemOpen
  \bibfield  {author} {\bibinfo {author} {\bibfnamefont {E.}~\bibnamefont
  {Montoya}}, \bibinfo {author} {\bibfnamefont {P.}~\bibnamefont {Omelchenko}},
  \bibinfo {author} {\bibfnamefont {C.}~\bibnamefont {Coutts}}, \bibinfo
  {author} {\bibfnamefont {N.~R.}\ \bibnamefont {Lee-Hone}}, \bibinfo {author}
  {\bibfnamefont {R.}~\bibnamefont {H\"ubner}}, \bibinfo {author}
  {\bibfnamefont {D.}~\bibnamefont {Broun}}, \bibinfo {author} {\bibfnamefont
  {B.}~\bibnamefont {Heinrich}}, \ and\ \bibinfo {author} {\bibfnamefont
  {E.}~\bibnamefont {Girt}},\ }\href {\doibase 10.1103/PhysRevB.94.054416}
  {\bibfield  {journal} {\bibinfo  {journal} {Phys. Rev. B}\ }\textbf {\bibinfo
  {volume} {94}},\ \bibinfo {pages} {054416} (\bibinfo {year}
  {2016})}\BibitemShut {NoStop}%
\end{thebibliography}%

\end{document}